\documentclass[12pt]{iopart}
\usepackage{graphicx}
\usepackage{pstricks}
\begin{document}
\jl{6}

\title{Zero-point length from string fluctuations}
\author{
Michele Fontanini\dag
\footnote{e-mail address: \texttt{fontanini@trieste.infn.it}},
Euro Spallucci\dag $\star$
\footnote{e-mail address: \texttt{spallucci@trieste.infn.it}}
T.Padmanabhan\ddag \footnote{e-mail address: \texttt{nabhan@iucaa.ernet.in}}	
}
\address{\dag\ Dipartimento di Fisica Teorica,
  Universit\`a degli Studi di Trieste,
         Strada Costiera 11, 34014 Trieste, Italy}
\address{\ddag\ IUCAA, 
Post Bag 4, Ganeshkhind, Pune - 411 007}
\address{$\star$\ Sezione INFN di Trieste, Trieste, Italy}

\begin{abstract}
One of the leading candidates for quantum gravity, viz. string theory, has the 
following features incorporated in it.
(i) The full spacetime is higher dimensional, with
(possibly) compact extra-dimensions;
(ii) There is a natural minimal length below which the concept
of  continuum spacetime needs to be modified by some deeper concept.
On the other hand, the
 existence of a minimal length (\textit{zero-point length}) in  
 four-dimensional spacetime, with obvious implications as UV regulator,
   has 
been often conjectured as a natural aftermath of any
correct  quantum  theory of gravity. 
We show that one can incorporate the apparently
unrelated pieces of information --- zero-point length, extra-dimensions,  
string $T$-duality --- in a consistent framework. This is done in terms of a
 modified Kaluza-Klein theory
that interpolates between (~high-energy~) string theory and (~low-energy~) 
quantum
field theory. In this model, the  zero-point length in four dimensions
 is a ``virtual memory'' of the length scale of compact extra-dimensions. 
 Such a scale turns out to be determined by $T$-duality inherited from the 
underlying
fundamental string theory. 
From a low energy perspective short distance infinities are cut off by
a minimal length which is proportional to the square root of the string
slope, i.e. $\sqrt{\alpha^\prime}$. Thus, we bridge the gap between the
 string theory domain and the low energy arena of point-particle
quantum field theory.
\end{abstract}

\submitted

\maketitle

There are several pieces of evidence that suggests that the three constants of 
Physics $c,\hbar$ and $G$
are likely to be replaced in a more fundamental theory by  three other constants 
$c,\hbar$ and $l_0^2$, where $l_0$ is a length scale and
$l_0^2$ has the dimensions of area. It has been conjectured for a long time 
that
a fundamental, zero-point, length of spacetime will
arise from the quantum gravity and --- in the naive models, 
$l_0^2\approx G\hbar/c^3$. [This idea has a long history: see for example,
\cite{snyder}, \cite{dewitt},\cite{yoneya},\cite{padma1} etc.; for a review 
and
more references, see \cite{garay} ]. The loop quantum gravity
(which works with four dimensions) has the same idea arising as a minimal area 
\cite{lqg} in terms of $l_0^2$. The existence of such a minimal area has 
immediate implications for the entropy of black holes and a holographic 
interpretation of gravity [see, e.g., \cite{tp1}]. It might even be possible 
see
the effects of such a minimal length in cosmological observations 
[for the earliest attempt, see \cite{tp2}; for a recent one see \cite{brand}].

This theoretical conjecture was taken forward in \cite{padma2} in which an 
unexpected connection between zero point length and a path integral duality 
was
discovered. It was shown that  if the path integral amplitude $\exp\left[\, i\, l\left(\, x, 
y\,\right)\,\right]$
used for the 
definition of the propagator 
is modified so that it is invariant under the duality
transformation $l\left(\, x\ ,y\,\right)\to l_0^2/l\left(\, x\ ,y\,\right)$,
then: (i) the propagator becomes UV finite, and (ii) $l_0$ represents a 
residual,
or zero-point length. This idea was followed up in
\cite{shanky} which showed that concrete computations can be performed in 
quantum
field theory leading to UV finite results. Given the fact
that string theory is a leading candidate for quantum gravity (which also has 
a
notion of T-duality) it is natural to ask whether these ideas can be combined 
in
a more formal manner.

In attempting this, we must remember that string theory introduces a 
fundamental
length scale as well as ---  a less evident, but not less important  ---  a 
second
length scale in the form of a \textit{compactification scale}. This is 
because,
in most models, the  extra-dimensions of  string target spacetime  must be 
compact
in order to be un-observable at the present day available energies.
In this letter we propose a single consistent framework for connecting the 
apparently unrelated pieces of information, i.e. zero-point length, 
extra-dimensions,  string $T$-duality.
  
We do this by introducing a low-energy $4D$  vacuum which keeps the
 memory of compact extra-dimensions only through topologically
 non-trivial fluctuations. These virtual
 processes are sensitive both to the presence of extra-dimensions and to
 the string excitation spectrum. This leads to a
 zero-point length is proportional to the compactification scale.
Furthermore, such  a scale respects the $T$-duality inherited from 
the underlying fundamental string theory. As the spectrum of
 closed strings cannot distinguish between a compactification radius $R$ and a radius 
$\alpha^\prime/R$
 no physical meaning can be attributed to length scale lower than 
 $\sqrt{\alpha^\prime}$.  We conclude that $l_0\propto \sqrt{\alpha^\prime}$.
  From a low energy perspective short distance infinities are cut off by
a minimal length, which is proportional to the square root of the string
slope defined as $\sqrt{\alpha^\prime}$. Thus, we bridge the gap between the
ultra-relativistic, ten-dimensional, string domain and the four-dimensional
low energy arena of point-particle quantum field theory.

The starting point of our technical analysis is the string Lagrangian
in the light-cone gauge \cite{book} ( for the sake of simplicity, we
shall consider only the case of a bosonic string )

\begin{equation}
\fl 
L= \frac{\pi\rho}{2}
\left(\,-2 \dot{x}^+_0 \dot{x}^- _0 + \dot x^i_0 \dot x_{0\, i}\,\right) 
+\frac{\rho}{2} \int_0^\pi  d\sigma
\left[\,  -2\dot{\eta}^+ \dot{\eta}^-  +\dot{\eta}_i\, \dot{\eta}^i  
+2\eta^{\prime\, +} \eta^{\prime\, -} 
 -{\eta}^\prime_i\, {\eta}^{\prime\, i}\,\right]\label{llc}
\end{equation} 
where, $x_0^\mu$ denotes the center of mass coordinate and $\eta^\mu$
the relative coordinate; $\rho=1/2\pi\alpha^\prime$ is the string tension;
 the index $i$ labels $d-1$ ``transverse'' space-like directions, for the sake
 of simplicity we shall consider the simplest case where only one of the 
transverse
 dimension is a circle of radius $R$. Given the Lagrangian, we can write
the transition
amplitude from an initial to a final configuration for the whole system as a 
path integral

\begin{equation}
\langle f\vert i\rangle \equiv \int_{x_{0,i}}^{x_{0,f}} [Dx_0]\, 
\int_{\eta_i}^{\eta_f} 
[D\eta] \: \exp \left[ i \int_0^T  d\tau\, 
L\left(\, \dot{x}_0\ , x_0\ ; \dot\eta \ , \eta \,\right) \right]
\label{start}
\end{equation}
where $x_{0,i}$ and $x_{0,f}$ represent initial and final position of the
string center of mass, while $\eta_i\equiv \eta(\, 0,\sigma)$ and
$\eta_f\equiv \eta(\, T,\sigma)$ are initial and final configuration
of the fluctuating part of the string.
In the low energy limit, described the quantum field theory, we will be 
interested
in the propagator for a particle-like object which will be described by the 
center of mass of the string. So, what we are really interested in, is the
 effective propagator for the string center of
mass propagating in vacuum in which all possible string fluctuations take 
place.
To account for virtual transitions among string states we need  to sum over
closed paths in the $\eta$ configurations space, i.e.,
 with the boundary condition $\eta_i=\eta_f$. This leads to the
 following expression

\begin{equation}
Z\left(\, T\,\right) \equiv \oint [D\eta] 
\: \exp \left[ i \int_0^T  d\tau\, L \left(\,\dot\eta \ , \eta\, \right) 
\right]
\label{zeta}
\end{equation}
where $L(\dot{\eta},\eta)$ represents the second term on the right hand
side of equation (\ref{llc}).
 Note that $Z$ does \textit{not} describe a physical gas of strings \cite{gas},
but a a mathematical quantity encoding the feature that  all kind of virtual 
transitions take place in the string physical vacuum.

In order to compute  $Z$ we first
 have to remove unphysical modes. This can be done by choosing  the
\textit{light-cone gauge}, which is a frame where
all the oscillations along $+$ direction are turned-off, i.e.
$\eta^+ =0\ ,\dot{\eta}^+ =0\ ,\eta^{\prime\, +}=0 $
Thus, in the light-cone  gauge the partition functional reads
 
\begin{equation}
Z\left(\, T\,\right)=\oint [\,D\vec\eta\, ] \: \exp \left[\, i\frac{\rho}{2} 
\int_0^T \int_0^\pi  d\tau\,  d\sigma \left(\,\dot{\vec{\eta}}^{\, 2} -
 \vec{\eta}^{\, \prime \, 2}\,\right)\, \right]
\label{zetalc}
\end{equation}
where only transverse physical oscillations are summed over. 
In analogy to the Coulomb gauge in electrodynamics, the light-cone
gauge allows to remove both
``timelike'' $\eta^+$ and ``longitudinal'' $\eta^-$ components of the
$\eta$-field.

Using the Fourier expansion for the transverse
coordinates we can  write $Z\left(\, T\,\right)$ in the form of a partition 
functional
 for two infinite families of transverse harmonic oscillators 

\begin{equation}
\fl
Z\left(\, T\,\right)=\oint \prod _{n=1} ^\infty  
\left[\, D\vec{x}_n\,\right]
\prod _{n=1} ^\infty  \left[\, D\tilde{\vec{x}}_n\,\right] 
\: \exp \bigg[i \frac{\rho \pi}{4} 
\int_0^T   d\tau\, \sum_{n=1}^\infty
\big[ (\dot{\vec{x}}^{\, 2}_n - 4n^2 \vec{x}^{\, 2}_n)
 +  ( \dot{ \tilde{\vec{x}}  }^{\, 2}_n -4n^2 \tilde{\vec{x}}^{\, 2}_n ) 
\big] \bigg] 
\label{Kos}
\end{equation}
Thus, the string partition function turns out to be an infinite product
of harmonic oscillator  path integrals computed over families of closed paths.

\begin{equation}
Z_{ho}\left(\, T\,\right)=\sum_{N,\tilde{N}=0}^\infty \exp 
\left[\,- i \, T \,
\left(\,N +\tilde{N}-\frac{d-1}{12}\,\right)\,\right]
\end{equation}
 But, whenever one (~ or more~) dimension is compact,  strings
can wrap around it an arbitrary number of times. Accordingly, we have to 
 take into account the contribution from the different winding
modes. For the sake of simplicity, let us consider again the case of a single
compact dimension. Then, we find

\begin{equation}
Z_{ho}\left(\, T\ , R\,\right)=
\sum_{N,\tilde{N},w =0}^\infty \exp \left[\, -i\, T \,\left(\, N +
\tilde{N}-\frac{d-1}{12} + w^2\,\frac{R^2}{\alpha ^\prime }\,\right)\,
\right]\label{zw}
\end{equation}
By including winding modes in Eq.~(\ref{zw}) we encode a topological feature
which makes the string substantially different from a pure ``gas'' of 
pointlike oscillators.

We can now  put all the results together and give a more definite meaning to 
the  center of mass kernel
in the vacuum which is filled up with both Kaluza-Klein type fluctuations
and the new kind of virtual processes brought
in by the string excitation modes:
 
\begin{eqnarray}
\fl
K\left(\, x_{0, f} -x_{0,i}\ ; T\,\right)&=&\sum_{N,\tilde{N},w =0}^\infty \exp
\left[\, -i T \,\left(\,N +
\tilde{N}-\frac{d-1}{12} + w^2\,\frac{R^2}{\alpha^\prime} \,\right)\,
\right] \times \nonumber \\
&&\int_{ x_0(0)= x_{0,i}  }^{ x_0(T)=x_ {0, f} } \left[\, Dx_0\,\right] 
\: \exp \left[ i \int_0^T  d\tau\, 
L\left(\, \dot{x}_0\ , x_0 \,\right) \right]
\label{kcm}
\end{eqnarray}
This path integral can be computed by weighting each 
path by its canonical action in phase-space :

\begin{equation}
\fl
S=\int_0^T \!  d\tau \left[ P_+ \dot{x}^+ \!+ P_- \dot{x}^- \!
+ P_j \dot{x}^j + P_d \dot{x}^d - \frac{1}{2 \pi \rho} \left( 2 P_+P_-  
+ P_j P^j +P_d^2 \right) \right]
\label{scan}
\end{equation}
Trajectories along the compact dimensions must satisfy periodic boundary 
conditions,
i.e. $x^d\left(\, T\,\right)=  x^d\left(\, 0\,\right)+ n\, l_0 $ where 
$l_0=2\pi\, R$.
Integration over center of mass degrees of freedom gives 
(see \cite{firstattempt}, \cite{emich} for details):
 
\begin{eqnarray}
\fl
K_{reg}\left(\, x_f-x_i\ ; T\,\right)=  
\left( \frac{1}{4 i\pi \alpha^\prime T}\right)^\frac{d-1}{2} 
\sum_{ N=0,w,n=1}^\infty \exp \left[
-\frac{ \left( x_f - x_i \right)^2 + n^2 l_0^2]}{4i
\alpha^\prime T} \right] \times \nonumber\\
 \exp\left[ - i T\left(2N + nw
 -\frac{d-1}{12} + \frac{w^2 R^2}{\alpha^\prime}\right)\right]
\label{Ksreg}
\end{eqnarray}
where, we have taken into account the level matching condition $\tilde{N}-N=nw 
$ and
 dropped out the zero-modes $n=0$ and $w=0$. The rationale
behind this subtraction is discussed in detail in \cite{firstattempt}, 
\cite{emich} and will not be repeated here. Further comments about this 
technical
step can be found at the end of this paper.
From (\ref{Ksreg}) it is possible to obtain the Green function by
integration over the unmeasurable lapse of time $T$ as follows:

\begin{eqnarray}
\fl
G\left( x_f-x_i \right) &\equiv& \left( 2\pi \right)^\frac{d-1}{2}
\int_0^\infty   dT \, e^{-i2\alpha^\prime m_0^2 T}\, 
K_{reg} \left( x_f - x_i ; \right) \nonumber\\
&=&\left(\frac{1}{2i\alpha^\prime}\right)^\frac{d-1}{2} 
\sum_{ N=0, w,n=1}^\infty \int_0^\infty   dT \, 
T^{-\frac{d-1}{2}} \exp\left[ -\frac{ \left( x_f - x_i \right)^2 + n^2 l_0^2}
{4i\alpha^\prime T} \right] \times \nonumber\\
&& \exp \left[ -i T\left(\, 2\alpha^\prime m_0^2 + 2N + nw
 -\frac{d-1}{12} + \frac{w^2 R^2}{\alpha^\prime}\,\right)\right]
\label{greall}
\end{eqnarray}
where the mass $m_0$ can be zero or non vanishing and has been introduced to 
account for low energy effects, e.g. spontaneous symmetry breaking.
In order to evaluate the short distance behavior of the Green function
(\ref{greall}) it is useful to Fourier transform it
\begin{equation}
G\left( p \right) =\sum_{ N=0\ ,\, w\ ,n=1}^\infty 
\frac{n l_0}{\sqrt{p^2 + M_{N,w,n}^2}}
\, K_1 \left( n l_0 \sqrt{p^2 + M_{N,w,n}^2}\right)
\label{grep}
\end{equation}
The mass term $M_{N,w,n}^2$ is defined as 
\begin{equation}
M_{N,w,n}^2 \equiv \frac{1}{\alpha^\prime} \left(  
 2N + nw -\frac{d-1}{2} + \frac{w^2 R^2}{\alpha^\prime} +
  2 \alpha^\prime m_0^2 \right)
\label{MNwn}
\end{equation}
At high energy( momentum ) the asymptotic behavior of the propagator 
(\ref{grep})
is essentially determined  by the lowest energy level $n=w=1$
\begin{equation}
G\left( p \right) \approx
\frac{ l_0}{\sqrt{p^2 + M_{0,1,1}^2}}
\, K_1 \left(  l_0 \sqrt{p^2 + M_{0,1,1}^2}\right)
\approx \frac{\sqrt{l_0}}{\left(\, p^2\,\right)^{3/4}}
\exp\left( -l_0\, \sqrt{p^2} \, \right)
\label{grepreg}
\end{equation}
As closed strings cannot probe compactification scales lower than 
$\sqrt{\alpha^\prime}$, then we can replace $l_0$ in 
(\ref{grepreg}) with $2\pi\sqrt{\alpha^\prime}$. Thus, it becomes manifest as 
UV divergences are exponentially suppressed at string energy scale.
Generalization to an hyper-torus with more than one compact dimensions is 
straightforward.
The standard 
Minkowski vacuum, with its pathological short-distance behavior, can be 
recovered in the limit $l_0\to 0$. 
It is worth observing that in our formulation the limit $l_0\to 0$ is 
equivalent to the infinite tension limit, i.e. $\alpha^\prime\to 0 $, where 
strings shrink to structureless points and the point-particle picture
of matter is recovered. 

Let us take stock of the result from a wider perspective. String theory uses 
$(4+D)$ dimensions of which $D$ are compact. The path integral in 
eq. (\ref{start}) represents the transition amplitude in the full theory. On 
the
other hand, the low energy quantum field theory uses only 4 dimensions and the
 theory is described by a propagator $G(x,y)$. To get $G(x,y)$ from the full 
 theory, it is appropriate to identify the center of mass of the string as 
 representing the particle of the quantum field theory. But then the 
propagator
 will be affected by the virtual fluctuations in the string vacuum. In 
particular,
 when $(x-y)^2$ is smaller than the size of the compact dimensions, these 
 fluctuations will lead to corrections to the propagator. These vacuum 
 fluctuations can be divided into two sets: the topologically trivial zero 
modes
 which do not probe the internal dimensions (the $n=0$ modes) and the 
 topologically nontrivial ones ($n\neq 0$). We have shown that when the latter 
 ones are retained, the following results are obtained: (a) The propagator 
picks
 up corrections which essentially involves replacing $(x-y)^2$ by 
$(x-y)^2+l_0^2$
 introducing a zero point length. (b) This is identical in form to the results 
 obtained earlier in \cite{padma2} and shows that the T-duality does lead to 
the
 path integral duality.(c) It provides a prescription for incorporating the 
 ``stringy" effects in the standard quantum field theory and the theory is now 
 UV-finite \cite{shanky}.

Once we select the
histories which are closed along the extra-dimensions for evaluating the path 
integral,  the full path 
integral factorizes into the product of the four dimensional path integral
for the propagator times the vacuum partition functional accounting for 
virtual
fluctuations along extra-dimensions. Our result arises from dropping
 the zero-modes, which describe topologically trivial 
fluctuations; i.e., fluctuations described by paths which can be 
continuously
shrunk to a point. With hindsight, it is clear how zero-modes bring 
ultraviolet divergences in, as they are ``blind'' to the extra-dimensions and
can probe arbitrary short distance.  The need of a ``by hand'' subtraction of 
the
zero-mode follows from the choice of the simplest toroidal compactification.
Hopefully, in a more sophisticated compactification scheme, or when we 
understand
the structure of some guiding principle behind the theory, the zero-mode will 
be
absent from the spectrum from the very beginning. This could be an effective
 criterion for selecting the appropriate kind of compact dimension(s) 
 among  many possible topologies. 

Thus, zero-point length in four dimensional spacetime can be seen as the 
\textit{virtual memory} of
the presence of compact extra-dimensions which can occur even much below the 
threshold energy needed to produce real Kaluza-Klein particles. 
Within the Kaluza-Klein quantum field 
theory picture, the actual value of $l_0$ remains undetermined. In the more
general framework provided by string theory, $T$-duality selects the 
\textit{unique} self-dual value for the compactification scale, and, 
accordingly determines $l_0=2\pi\sqrt{\alpha^\prime}$.
One could, therefore, expect to see deviations from the
theoretical predictions of standard quantum field theory due to the presence
of the modified Feynman propagator (\ref{grep}) at an intermediate energy 
regime
much below the string scale. In this respect, 
the most optimistic scenario is offered by the $TeV$ scale unification
models, where the string scale is lowered down to a few $TeV$ \cite{tev},
\cite{koko}.  Such a ``low-energy'' unification can be realized provided the 
extra-dimensions are compactified to a ``large'' radius of some fraction of 
millimeter. In this case $l_0\approx 10^{-17}cm.$ and its presence would be 
detectable in the high-energy scattering experiments \cite{sabine} planned for 
the next generation of particle accelerators. 

One of the authors, (E.S.), thanks  T. Padmanabhan
for triggering his interest to find out a link between zero-point
length and string theory \cite{firstattempt};  the same author would also like 
to
thank  S. Shankaranarayanan for useful discussions.

\end{document}